\documentclass[aps,prl,twocolumn,showpacs,superscriptaddress,groupedaddress]{revtex4-1}  
\usepackage{graphicx}  
\usepackage{dcolumn}   
\usepackage{bm}        
\usepackage{amssymb}   
\usepackage[utf8]{inputenc}

\newcommand{\ket}[1]{\left| #1 \right>} 

\begin{document}

\title{Predictability of measurements}
\author  {Bálint Szabó}
\affiliation{Department of Biological Physics, Eötvös University, Pázmány Péter stny 1/A, Budapest, H-1117, Hungary.}
						
\date{\today}

\begin{abstract}
The second law of thermodynamics states that entropy increases (or does not change) by time in an isolated system. As microscopic physical laws are reversible, the origin of irreversibility is not straightforward. Although the outcome of a measurement on a pure quantum state is not fully predictable due to the Heisenberg uncertainty principle, quantum and finite entropy uncertainties are thought to be fundamentally different. We propose to calculate the predictability of measurements comprising both quantum and entropic uncertainties. We show that the unpredictability of measurements is identical to entropy in case of semiclassical statistical mechanics, and it increases by time in a pure entangled quantum state as a result of quantum measurement.
\end{abstract}


\maketitle



Entropy, a central quantity of statistical physics calculated from the probability of possible states is a widely accepted measure of information content. It is also considered to be a measure of disorder or rather the amount of information unavailable for the theory describing the system. Thermodynamic definition of entropy is ascribed to Clausius. Statistical explanation of entropy based on the microscopic components of a physical system was laid down by Boltzmann. von Neumann defined the entropy of quantum systems by changing classical microscopic states to quantum states resulting in zero for pure states. In information theory a similar measure to physical entropy was introduced by Shannon. Throughout this paper, for the sake of simplicity, we calculate entropy with logarithm to base $2$ and without the Boltzmann constant. Instead of the probabilistic definition of entropy, algorithmic randomness measuring the information content of individual microstates was proposed to be applied in physics by Zurek \cite{algorithmic_randomness} .


Second law of thermodynamics stemming from the early work of Carnot and Clausius states that entropy increases (or does not change) by time in an isolated system. This law breaks the symmetry of time. As microscopic physical laws are reversible, the origin of irreversibility is not straightforward. (See \cite{q_second_law_TD} and references therein.) 


Quantum measurement according to von Neumann can be described by a projection in the Hilbert space to an eigenstate of the operator representing the quantity being measured. The measurement problem, i.e., how the projection happens is still not fully resolved. Studies of quantum decoherence attempt to give a probabilistic explanation \cite{decoherence}. The measuring process itself is not discussed in this paper.

The famous EPR paradox \cite{EPR} introduced by Einstein, Podolsky and Rosen points out the peculiar feature of quantum measurement on entangled particles. Bohm and Aharonov gave both theoretically and experimentally simpler example of the EPR paradox \cite{EPR_Bohm_Aharonov} with two spin-$1/2$ particles in a singlet state.  This state is isotropic, i.e., measuring the spin of one particle along an arbitrary direction infers opposite spin for the other particle along the same direction.
Considering the spin and the spatial part of the single particle states, the initial two-particle state of the EPR paradox can be described by the following antisymmetric ket vector in the center-of-momentum frame:

\begin{equation}\label{initial_EPR_state}
\ket{\Psi_i}=1/2(\ket{\alpha}\ket{\beta}-\ket{\beta}\ket{\alpha})(\ket{+p}\ket{-p}+\ket{-p}\ket{+p}),
\end{equation}

where $\ket{\alpha}$ and $\ket{\beta}$ are the Zeeman eigenstates of the spin operator ($I_z$) with $+1/2$ and $-1/2$ eigenvalues; $(\ket{+p}$ and $(\ket{-p}$ are representing the spatial parts of the single particle states with momentums $+p$ and $-p$, respectively. Assuming that one of the distant observers measures $I_z$ of the particle moving towards her with a momentum of $+p$, and finds $+1/2$, the final, still antisymmetric state of the two-particle system, according to her, will be as follows:

\begin{equation}
\ket{\Psi_f}=1/\sqrt{2}(\ket{\alpha}\ket{\beta}\ket{+p}\ket{-p}-\ket{\beta}\ket{\alpha})\ket{-p}\ket{+p}),
\end{equation}

Bell showed that the introduction of further hypothetical, so called hidden variables to amend the description given by quantum mechanics cannot solve the EPR paradox \cite{Bell}. According to Bell one should make do with the predictions of quantum mechanics with no hope of a more complete theory.


In physics there is to be a tight correspondence between the abstract objects of theories and observable quantities (shortly observables). The correspondence, however, is never perfect. Both in statistical and quantum mechanics the output of measurements can be predicted only with some uncertainty due to finite temperature and Heisenberg uncertainty, respectively. These are usually taken into account by introducing noise, i.e., random parameters into the theoretical models.

The concept that entropy is calculated from the probabilities of possible theoretical states (mathematical models) is very reasonable if all the observables can be deduced from the model. In quantum systems it is not the case according to the Copenhagen interpretation devised by Bohr and Heisenberg.

We propose a more pragmatic definition to measure the uncertainty content of a system: the amount of observable information hidden from the environment. This is closely related to the predictability of possible measurements. To calculate the unpredictability ($P$) of a measurement we need to count the possible outcomes (with equal probabilities) and convert this number to a logarithmic scale as we claim additivity for independent systems:

\begin{equation}\label{P}
P=log_2{N},
\end{equation}

where $N$ denotes the number of possible outcomes of the measurement. If the outcomes are not equally probable, we need to use the formula:

\begin{equation} \label{P_diff_probabilities}
P=-\sum_{i}{p_{i}log_2{p_{i}}},
\end{equation}

where $p_{i}$ denotes the probability of the $i ^{th}$ outcome. If more than one independent observables can be measured, one should consider all the possible outcomes with probabilities $p_{ij}$:

\begin{equation} \label{P_multiple_observables}
P=-\sum_{i,j}{p_{ij}log_2{p_{ij}}},
\end{equation}

where $i$ indexes the outcome of the $j^{th}$ independent observable. Independent means that there is no theoretical correlation between these observables, e.g., the $x$ and $y$ components of the spin ($I_x$ and $I_y$) of a spin-$1/2$ particle are independent in the $\ket{\alpha}$ state. It has to be emphasized that not all the independent observables can be measured simultaneously due to the Heisenberg uncertainty principle. Nevertheless, the formula above still can be reasonable.

In semiclassical statistical mechanics each microscopic state of the system corresponds to a set of possible values of observables, e.g., the momentums and coordinates of particles. Knowing the state means that one knows the values of observables. This relation leads to the equality between unpredictability of measurements and entropy. In case of a single observable using Eq. (\ref{P}) or (\ref{P_diff_probabilities}) this is trivial. In case of multiple independent observables the probability of the $k^{th}$ state, $X_k$ is the product of probabilities of the values of observables defining the state:

\begin{equation}\label{P(X_k)-p_alpha,j}
p(X_k)=\prod_{j}{p_{\alpha(k,j),j}},
\end{equation}

where $\alpha(k,j)$, a function of $k$ and $j$ indexes the value of the $j^{th}$ observable in the $X_k$ state. The sum of the probabilities of those states that implicate the $i^{th}$ outcome of the $j^{th}$ observable results in the probability $p_{ij}$ introduced in Eq. (\ref{P_multiple_observables}):

\begin{equation} \label{p_ij-X_k}
p_{ij}=\sum_{k}{p(X_k)\delta_{i,\alpha(k,j)}},
\end{equation}

where $\delta$ stands for the Kronecker delta. Entropy of the system:

\begin{equation}\label{S-X_k}
S=-\sum_k{p(X_k)log_2{p(X_k)}}.
\end{equation}

Equations (\ref{P_multiple_observables}), (\ref{P(X_k)-p_alpha,j}), (\ref{p_ij-X_k}), and (\ref{S-X_k}) deduce

\begin{equation}
S=P.
\end{equation}

Unpredictability of measurements on a single spin-$1/2$ in a pure quantum state is $2$, as two components of the spin are uncertain with $2$-$2$ possible outcomes and $1/2$ probabilities. For instance in the  $\ket{\alpha}$ state the measurement of either $I_x$ or $I_y$ has two equally likely outcomes. The von Neumann entropy of this state is $0$. In a mixed state of $\ket{\alpha}$ and $\ket{\beta}$ with $1/2$-$1/2$ probabilities the von Neumann entropy is $1$. In this state all three components of the spin are totally uncertain. Accordingly, unpredictability results in $3$. The difference in the information content of the pure and mixed states is equally $1$ calculated either using the von Neumann entropy or unpredictability of measurements.

Measuring the spin of a single spin-$1/2$ particle in the $\ket{\alpha}$ state along an arbitrary axis will not change the predictability of the system, as it remains $2$. The von Neumann entropy is also unchanged.

However, unpredictability increases in an EPR experiment after the measurement. In the initial state described by Eq. (\ref{initial_EPR_state}) it is $3$ as the number of independent observables with $2$-$2$ possible outcomes and $1/2$ probabilities is $3$. Although all three components of the single particle spins are uncertain, spin components of one particle can be deduced from the values of the other particle. In other words, the total number of uncertain observables is $6$ but there are also $3$ correlations between them.  In the final state unpredictability has been increased to $4$ with $4$ independent uncertain observables: $I_x$ and $I_y$ of both particles. The increase in unpredictability is accompanied by the loss of correlation between these observables.

If the two measuring apparatus of the EPR experiment is modeled by two additional spin-$1/2$ particles in an arbitrary initial state and in a final state similar to the final state of the measured particle we can calculate the change in predictability for the full system including the measuring particles. Measuring particles are considered to be not entangled with the EPR particles (singlet pair of particles) before and after the measurement. As the predictability of the measuring single particles are unchanged we find that the predictability of the full system also decreases due to the measurement.

In case of observables described by continuous variables unpredictability can be infinite assuming an infinitely precise measurement. As in reality measurements are not infinitely precise, the change in unpredictability is finite even in case of continuous physical quantities.

Although an abstract quantum state may be the very best model of a physical system described by an observer according to Bell \cite{Bell}, it may still not be identical to reality. Abstract models are constructed by observers on the basis of information relating the physical system observed. Some observers may have less information than others. Entropy may be attributed to conflicting models compatible with different (possible) observers. A quantum system before and after the measurement is described by different states that cannot be merged into one coherent time-dependent quantum state. Similar applies to the models constructed by the two distant observers in the EPR experiment. One of them already has $\ket{\Psi_f}$ in mind, while the other is unable to update $\ket{\Psi_i}$ lacking the information of the measurement. 

Theory of Relativity states that the direction of information flow cannot be space-like, i.e., a message can be transferred only to the future not to the past. Consequently a process accompanying the flow of information is also time-like. If the measurement process, i.e., the information flow from the measured system to the measuring instrument is accompanied by the increase of unpredictability, it may lead to the second law of thermodynamics.



\begin{thebibliography}{6}

\bibitem{algorithmic_randomness}

W. H. Zurek, Phys. Rev. A {\bf 40}, 4731 (1989).

\bibitem{q_second_law_TD}
    
J. Gemmer, A. Otte, G. Mahler, Phys. Rev. Lett. {\bf 86}, 1927 (2001).
 
\bibitem{decoherence}
 
M. Schlosshauer, Decoherence and the quantum-to-classical transition, Springer (2007).

\bibitem{EPR}

A. Einstein, B. Podolsky, N. Rosen, Phys. Rev. {\bf 47}, 777 (1935).

\bibitem{EPR_Bohm_Aharonov}

D. Bohm, Y. Aharonov, Phys. Rev. {\bf 108}, 1070 (1957).

\bibitem{Bell}

J. S. Bell, Physics, {\bf 1}, 195 (1964).

\end{thebibliography}
\end{document}